# Investigation of particle distributions in Xe-Xe collision at $\sqrt{s_{NN}} = 5.44$ TeV with Tsallis statistics

Hai-Fu Zhao, Bao-Chun Li and Hong-Wei Dong

College of Physics and Electronics Engineering, Collaborative Innovation Center of Extreme Optics, State Key Laboratory of Quantum Optics and Quantum Optics Devices, Shanxi University, Taiyuan 030006, China

**Abstract:** The distribution characteristic of final-state particles is one of significant parts in high energy nuclear collisions. The transverse momentum distribution of charged particles carries essential evolution information about the collision system. Tsallis statistics is used to investigate the transverse momentum distribution of charged particles produced in Xe-Xe collisions at $\sqrt{s_{NN}} = 5.44$ TeV. On the basis, we reproduce the nuclear modification factor of the charged particles. The calculated results agree approximately with the experimental data measured by the ALICE Collaboration.

## 1. Introduction

One of the major goals of high energy nucleus-nucleus (AA) collisions is to study quark-gluon plasma (QGP) at high energy density and high temperature. The Large Hadron Collider (LHC) has performed different species of collisions at one or more energies, such as lead-lead, proton-lead and proton-proton collisions. The Xe-Xe ion collision [1, 2] at $\sqrt{s_{NN}} = 5.44$ TeV is a new collision experiment and is an intermediate-size collision system at the LHC. Since the mass number value of the xenon is between the proton and the lead, it helps us to understand the system-scale effect of the final-state particle properties in ion collisions at high energy [3-6]. Many charged particles are produced and measured in the AA collisions. The investigation of the particle spectra is of great interest and is very helpful for comprehending the collision reaction mechanism and the particle production process in the different species of collision systems at different center-of-mass energies [7-13].

With respect to the final-state observations, the experimental transverse momentum $p_T$ spectrum is of great significance in understanding the production process of the moving particles. In past years, theoretical efforts have been carried out in statistical models to analyze the particle spectra over a broad range of collision energies [14-18]. At RHIC and LHC energies, the $p_T$ spectra have been investigated intensively in various collision systems like Au+Au, Pb+Pb and pp at different energies. A statistical model can achieve some features in treating the multiparticle system in the RHIC and the LHC. Recently, the ALICE Collaboration reported the $p_T$ spectra and nuclear modification factors of charged particles produced in Xe–Xe collisions at $\sqrt{s_{NN}} = 5.44$ TeV [1]. The nuclear modification factor $R_{AA}$ is also an important observation and can provide information about the dynamics of QGP matter at extreme densities and temperatures [19-26].

In this paper, we discuss the $p_T$ spectra and the nuclear modification factor $R_{AA}$ in Tsallis statistics. By the investigation of the $p_T$ spectra, we extract the parameters, which provide the calculation foundation for the nuclear modification factor $R_{AA}$.

## 2. Description of the particle distribution in Tsallis statistics

Tsallis statistics has been widely used to study the properties of final-state particles produced in nucleus-nucleus and proton- proton collisions at high energy [27-30]. According to Tsallis statistics, the number of the final-state particles is

$$N = gV \int \frac{d^3p}{(2\pi)^3} \left[1 + (q-1)\frac{E-\mu}{T}\right]^{-\frac{1}{q-1}}, \quad (1)$$

where $g$ and $\mu$ are a degeneracy factor and a chemical potential of the multiparticle system, respectively. The $T$ and $q$ are a Tsallis temperature and the degree parameter of deviation from equilibrium, respectively. At $\mu = 0$, the transverse momentum distribution is

$$\frac{d^2N}{dydp_T} = \frac{gVp_T\sqrt{p_{p_T}^2+m^2}\cosh y}{(2\pi)^2}\left[1+(q-1)\frac{\sqrt{p_{p_T}^2+m^2}\cosh y}{T}\right]^{-\frac{1}{q-1}}. \quad (2)$$

The nuclear modification factor $R_{AA}$ acts as a probe to understand the nuclear medium effect in the AA collision and is a measure of the particle production modification. It is typically expressed as a ratio of the particle $p_T$ spectra in AA collisions to that in pp collisions, *i. e.*

$$R_{AA}(p_T) = \frac{d^2N^{AA}/dydp_T}{\langle T_{AA}\rangle d^2\sigma^{pp}/dydp_T}, \quad (3)$$

where $N^{AA}$ is the production yield in AA collisions and $\sigma^{pp}$ is the production cross section in pp collision. The average nuclear overlap function $\langle T_{AA}\rangle$ is estimated via a Glauber model of nuclear collisions. The $R_{AA}$ is also expressed as

$$R_{AA} = \frac{f_{fin}}{f_{in}}, \quad (4)$$

where $f_{in}$ is the distribution of the initial particles produced at an early time of the hadronization. Then, these particles interact with the medium system. The function $f_{fin}$ is the distribution final-state particles, which no longer interact with each other.

According to Boltzmann transport equation, the distribution of particles $f(x, p, t)$ is

$$\frac{df(x, p, t)}{dt} = \frac{\partial f}{\partial t} + v\cdot\nabla_x f + F\cdot\nabla_p f = C[f]. \quad (5)$$

The evolution of the particle distribution is attributed to its interaction with the medium particles. The term $v$ and $F$ are a velocity and an external force, respectively. In relaxation time approximation, the collision term $C[f]$ is given by

$$C[f] = -\frac{f-f_{eq}}{\tau}, \quad (6)$$

where $\tau$ is the relaxation time. The Boltzmann local equilibrium distribution $f_{eq}$ is

$$f_{eq} = \frac{gV}{(2\pi)^2} p_T m_T e^{-\frac{m_T}{T_{eq}}}, \quad (7)$$

where $T_{eq}$ is the equilibrium temperature of the QCD phase transition. Considering $\nabla_x f = 0$ and $F = 0$, the distribution of particles $f(x, p, t)$ is

$$\frac{df(x,p,t)}{dt} = \frac{\partial f}{\partial t} = \frac{f - f_{eq}}{\tau}. \tag{8}$$

A solutions of the equation is

$$f_{fin} = f_{eq} + (f_{in} - f_{eq})e^{-\frac{t_f}{\tau}}, \tag{9}$$

where $t_f$ is the freeze-out time. The initial distribution is taken as the Tsallis distribution, *i. e.* Equation (2). Therefore, the final-state distribution is

$$f_{fin} = \frac{gV}{(2\pi)^2} p_T m_T e^{-\frac{m_T}{T_{eq}}} + \frac{gV}{(2\pi)^2} p_T m_T \left\{ \left[1 + (q-1)\frac{m_T}{T}\right]^{-\frac{q}{q-1}} - e^{-\frac{m_T}{T_{eq}}} \right\} \cdot e^{-\frac{t_f}{\tau}}. \tag{10}$$

Then, the nuclear modification factor $R_{AA}$ is obtained as,

$$\begin{aligned} R_{AA} &= \frac{f_{eq}}{f_{in}} + \left(1 - \frac{f_{eq}}{f_{in}}\right) e^{-\frac{t_f}{\tau}} \\ &= \frac{e^{-\frac{m_T}{T_{eq}}}}{\left(1 + (q-1)\frac{m_T}{T}\right)^{-\frac{q}{q-1}}} + \left[1 - \frac{e^{-\frac{m_T}{T_{eq}}}}{\left(1 + (q-1)\frac{m_T}{T}\right)^{-\frac{q}{q-1}}}\right] e^{-\frac{t_f}{\tau}}. \end{aligned} \tag{11}$$

The equation is the calculation basis of the nuclear modification factor. In the relaxation time approximation, the $R_{AA}$ is derived in the Tsallis statistics.

## 3. Discussions and conclusions

In the section, we discuss the transverse momentum spectra and the nuclear modification factor of charged particles produced in Xe-Xe collisions at $\sqrt{s_{NN}} = 5.44$ TeV. The transverse momentum contributes significantly to the characterization of the matter formed in high energy collisions because $p_T$ is sensitive to the matter properties at an early time. The transverse momentum spectra in the kinematic range $0.15 < p_T < 50$ GeV/c and |η| < 0.8 are presented for

nine centrality classes in Fig. 1. The filled circles indicate the experimental data measured by the ALICE Collaboration [1]. The lines are the results of the Eq. (2). The value of $T_{eq}$ is 0.24 GeV. The model results are in agreement with the experimental data. The maximum value of $\chi^2$ is 0.942 and the minimum is 0.205. The other parameters used in the calculation are listed in Table 1. The nonequilibrium degree $q$ is a constant value. The freeze-out time $T$ increases with increasing collision centrality. The final-state transverse momentum spectra for different centralities are determined by the temperature $T$, at which there are no interactions between the final-state particles. By the analysis of the $p_T$ spectra, the thermodynamics parameters are extracted.

The nuclear modification factor is also an important observation and is a measure of the particle-production modification. In Fig. 1, we compare the $p_T$ spectra of the model results and the experiment data, and can extract the parameters, which are required in the calculation of the nuclear modification factor $R_{AA}$. Fig. 2 presents the nuclear modification factor $R_{AA}$ of charged particles as a function of $p_T$ in Xe-Xe at $\sqrt{s_{NN}} = 5.44$ TeV collisions. The filled circles indicate the experimental data measured by the ALICE Collaboration [1]. The lines are the results of the Eq. (11). The parameters used in the calculation are determined by the model results in Fig. 1. The nuclear modification factor $R_{AA}$ depends strongly on the collision centrality. The $R_{AA}$ rises linearly at low $p_T$ (about below 2.2 GeV). At high $p_T$, the $R_{AA}$ first declines linearly and then rises slowly. The model can approximately describe the nuclear modification factor at the high $p_T$ region.

Both experimentally and theoretically, the study of the particle spectra can contribute to our understanding of the particle production and the evolution dynamics in the collision system. Tsallis statistics has attracted extensive attention due to the investigation of final-state particles produced in nuclear collisions at high energies. In our previous work [31-34], the statistics model is only used to study the transverse momentum spectra of particles produced in nuclear collisions at different energies. It is successful in explaining the experimental data of the transverse momentum spectra and can obtain some thermodynamics information, such as the temperature, the chemical potential, and so on. The present work is a new attempt. The model is improved by Tsallis statistics in relaxation time approximation. Considering relaxation time approximation of the collision term, we achieve the final-state distribution by solving Boltzmann transport equation, where the initial distribution is inserted consistently. And, the expression of the $R_{AA}$ calculation in Tsallis statistics is derived. The improved model can not only describe transverse momentum spectra, but also reproduce nuclear modification factor of particles in Xe-Xe collisions at $\sqrt{s_{NN}} = 5.44$ TeV in different centrality classes.

## Acknowledgments

This work was supported by National Natural Science Foundation of China under Contract No. 11575103, Shanxi Provincial Natural Science Foundation under Grant No. 201701D121005, and Scientific and Technological Innovation Programs of Higher Education Institutions in Shanxi (STIP) Grant No. 201802017.

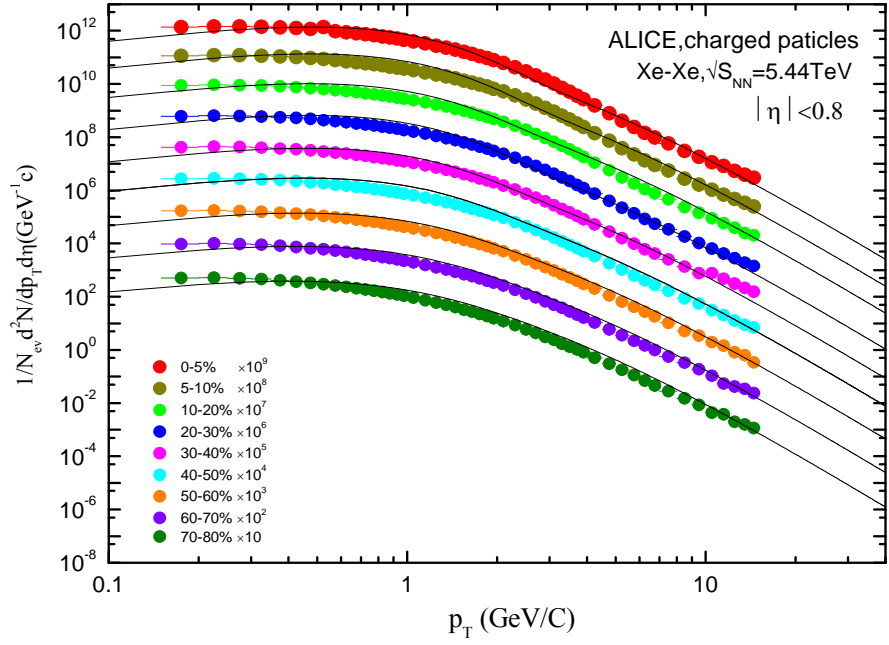

Figure 1: Transverse momentum distributions of charged particles produced in Xe-Xe collision at $\sqrt{s_{NN}} = 5.44$ TeV. The filled circles indicate the experimental data in 0-5%, 5-10%, 10-20%, 20-30%, 30-40%, 40-50%, 50-60%, 60-70% and 70-80% centrality classes [1]. The lines are the results of the Eq. (2).

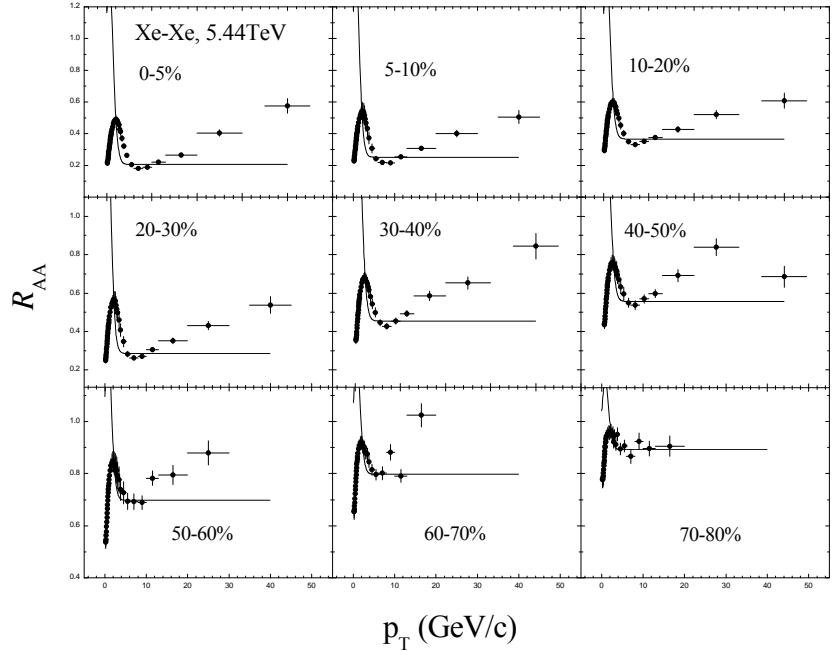

Figure 2: Nuclear modification factor $R_{AA}$ as a function of $p_T$ in Xe-Xe collision at $\sqrt{s_{NN}} = 5.44$ TeV. The filled circles indicate the experimental data in 0-5%, 5-10%, 10-20%, 20-30%, 30-40%, 40-50%, 50-60%, 60-70% and 70-80% centrality classes [1]. The lines are the results of the Eq. (11).

Table 1. Values of $q, T$ and $t$ taken in Figure 1.

| Centrality | $q$ | $T$ | $t_f/\tau$ |
|:---:|:---:|:---:|:---:|
| 0-5% | 1.125 | 0.196 | 1.581 |
| 5-10% | 1.125 | 0.191 | 1.381 |
| 10-20% | 1.125 | 0.187 | 1.005 |
| 20-30% | 1.125 | 0.185 | 1.252 |
| 30-40% | 1.125 | 0.180 | 0.788 |
| 40-50% | 1.125 | 0.178 | 0.586 |
| 50-60% | 1.125 | 0.175 | 0.360 |
| 60-70% | 1.125 | 0.169 | 0.226 |
| 70-80% | 1.125 | 0.165 | 0.115 |